\documentclass{PoS}
\usepackage{amsfonts}
\usepackage{mathrsfs}
\usepackage[ansinew]{inputenc}      
\usepackage{amssymb}                
\usepackage{amsmath}                
\usepackage[]{graphicx}
\usepackage{mathrsfs}
\usepackage{verbatim}
\usepackage{overpic}
\usepackage{xspace}
\usepackage{lineno}
\usepackage{enumitem}
\usepackage[font=small,skip=3mm]{caption}
\usepackage{setspace}
\usepackage{array}
\usepackage{lscape}
\usepackage{rotating}

\newcommand{\ttbar}{\ensuremath{t\bar{t}}\xspace}
\newcommand{\etmiss}{\ensuremath{E \kern-0.6em\slash_{\rm T}}\xspace}
\newcommand{\etmissx}{\ensuremath{E \kern-0.6em\slash_{\rm x}}\xspace}
\newcommand{\etmissy}{\ensuremath{E \kern-0.6em\slash_{\rm y}}\xspace}

\newcommand{\mlb}{\ensuremath{m_{\ell b}}\xspace}

\newcommand{\ljets}{\ensuremath{\ell\!+\!{\rm jets}}\xspace}
\newcommand{\dilep}{\ensuremath{\ell\ell}\xspace}

\newcommand{\etal}{\textit{et~al.}\xspace}

\newcommand{\GeV}{\ensuremath{\textnormal{GeV}}\xspace}
\newcommand{\TeV}{\ensuremath{\textnormal{TeV}}\xspace}

\newcommand{\dif}{\ensuremath{{\rm d}}}

\newcommand{\met}{\ensuremath{E_\mathrm{T}^\mathrm{miss}}\xspace}

\newcommand{\fb}{\ensuremath{{\rm fb}^{-1}}\xspace}

\newcommand{\mt}{\ensuremath{m_t}\xspace}
\newcommand{\mtpole}{\ensuremath{m_t^{\rm pole}}\xspace}

\newcommand{\mtgen}{\ensuremath{m_t^{\rm gen}}\xspace}

\newcommand{\rbq}{\ensuremath{R_{bq}}\xspace}

\newcommand{\pt}{\ensuremath{p_{\rm T}}\xspace}

\newcommand{\stt}{\ensuremath{\sigma_{t\bar t}}\xspace}

\newcommand{\sseven}{\ensuremath{\sqrt s=7~\TeV}\xspace}
\newcommand{\seight}{\ensuremath{\sqrt s=8~\TeV}\xspace}

\newcommand{\sttj}{\ensuremath{\sigma_{t\bar t+1~{\rm jet}}}\xspace}

\newcommand{\stat}{\ensuremath{{\rm(stat)}}\xspace}
\newcommand{\syst}{\ensuremath{{\rm(syst)}}\xspace}

\title{Measurements of the top quark mass\\
with the ATLAS detector}

\ShortTitle{Top quark mass from ATLAS}

\author{\speaker{Oleg Brandt on behalf of the ATLAS Collaboration}\\
        Kirchhoff-Institut f\"ur Physik, Im Neuenheimer Feld 227,\\
        69120 Heidelberg, Germany\\
        E-mail: \email{oleg.brandt@kip.uni-heidelberg.de}}

\abstract{
The top quark mass is one of the fundamental parameters of the Standard Model. In these proceedings, recent measurements of the top quark mass in $pp$ collisions at centre-of-mass energies of $\sqrt s=7$ and 8~TeV data in Run I of the Large Hadron Collider using the ATLAS detector are reviewed. A measurement using lepton+jets events is presented, where a multidimensional template fit is used to constrain the uncertainties on the energy measurements of jets. The measurement is combined with a measurement using dilepton events. In addition, novel measurements aiming to measure the mass in a well-defined scheme are presented. These measurements use precision theoretical QCD calculations for both inclusive $\ttbar$ production and ttbar production with an additional jet to extract the top quark mass in the pole mass
scheme.

}

\FullConference{The European Physical Society Conference on High Energy Physics\\
		22--29 July 2015\\
		Vienna, Austria}

\begin{document}

\section{Introduction}
Since its discovery~\cite{bib:discoverydzero,bib:discoverycdf}, the determination of the top quark mass \mt, a fundamental parameter of the standard model (SM), has been one of the main goals of the CERN Large Hadron Collider (LHC) and of the Fermilab Tevatron Collider. Indeed, \mt and the masses of the $W$ and Higgs bosons are related through radiative corrections that provide an internal consistency check of the SM~\cite{bib:lepewwg}. Furthermore, \mt dominantly affects the stability of the SM Higgs potential
~\cite{bib:vstab1,bib:vstab2,bib:vstab3}.
Currently, with $\mt=173.34\pm0.76~\GeV$, a world-average combined precision of about 0.5\% has been achieved~\cite{bib:combiworld}.
Measurements of properties of the top quark other than \mt\ at ATLAS~\cite{bib:atlas} are reviewed in Refs.~\cite{bib:proptalk,bib:topresatlas}. 
%
At the LHC, top quarks are mostly produced in pairs via the strong interaction.
In the SM, the top quark decays to a $W$~boson and a $b$~quark nearly 100\% of the time, resulting in $\ttbar\to W^+W^-b\bar b$. 
Thus, $\ttbar$ events are classified according to $W$ boson decays as ``dileptonic''~(\dilep), ``lepton+jets'' (\ljets), or ``all--jets''. Single top production in the dominant $t$-channel proceeds predominantly through the $qg\to q't\bar b$ process.

\section{Top quark mass in single top events}\label{sec:st} 

\begin{figure}
\centering
\begin{overpic}[clip,height=4.5cm]{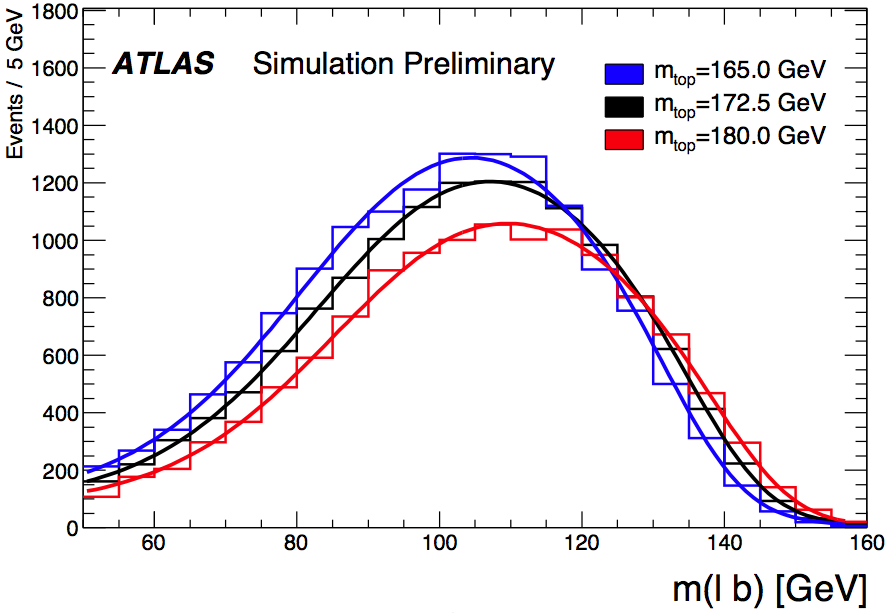}
\put(15,48){(a)}
\end{overpic}
\begin{overpic}[clip,height=4.5cm]{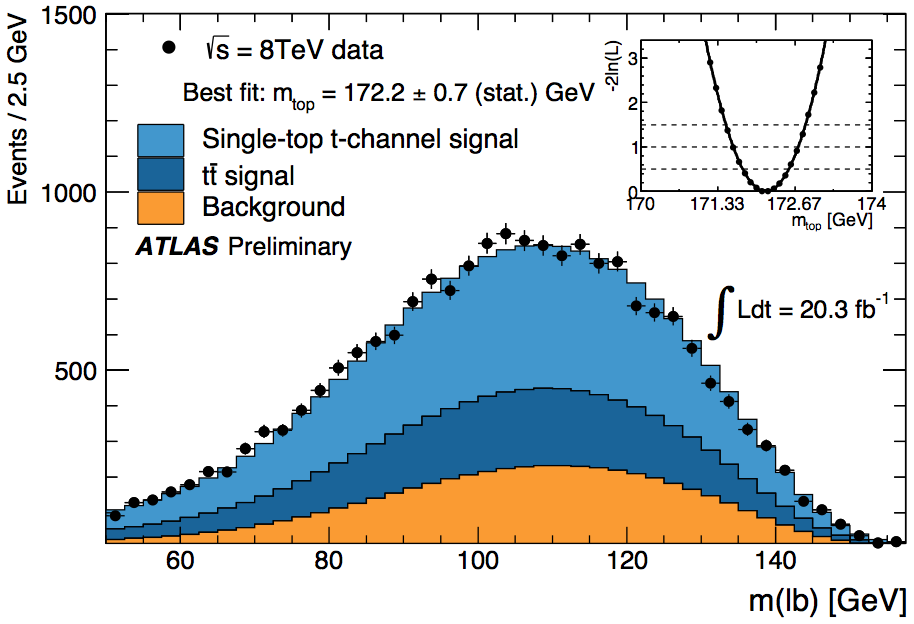}
\put(15,33){(b)}
\end{overpic}
\caption{
\label{fig:st}
{\bf(a)} Dependence of the \mlb distribution of all processes involving top quarks on \mt using simulated MC samples, together with the fitted signal probability densities~\cite{bib:st}. 
{\bf(b)} The distribution in $\mlb$ in topologies enriched with single top events using 20.3~\fb of data at \seight~\cite{bib:st}. The templates for the single top and \ttbar predictions are scaled to the best-fit values in data, alongside with the background template. 
}
\end{figure}

The first measurement of \mt in topologies enriched with $t$-channel single top events is performed using 20.3~\fb of data at \seight~\cite{bib:st}. This analysis is complementary to traditional approaches using \ttbar events due to factors like background composition or colour-connections, resulting in a different impact of systematic uncertainties. The selection requires a single isolated lepton ($e$ or $\mu$) with $\pt>25$~GeV and $\geq2$ jets with one $b$-tag and $\pt>30~\GeV$ within the well-instrumented region of the detector. Jets with $2.75<|\eta|<3.5$ and  $\pt>35$ are accepted as one forward  jet is expected in $qg\to q't\bar b$. Missing transverse momentum $\met>30~\GeV$ and topo\-logical requirements are imposed. The purity is increased to $\approx$50\% through a neural network constructed from 12 input variables.
The extraction of \mt is performed with the template method, i.e., by fitting the data using template distributions obtained from MC and parametrised in \mtgen. The \mlb observable chosen for the templates is shown in Fig.~\ref{fig:st}(a). The best fit to data is given in Fig.~\ref{fig:st}(b) and provides $\mt=172.2\pm0.7~\stat\pm2.0~\syst~\GeV$. Systematic uncertainties are dominated by the jet energy scale (JES) (1.5~\GeV) 
as well as the parton showering (PS) and hadronisation (0.7~\GeV).

\section{All-hadronic channel} \label{sec:jj}

\begin{figure}
\centering
\begin{overpic}[clip,height=5cm]{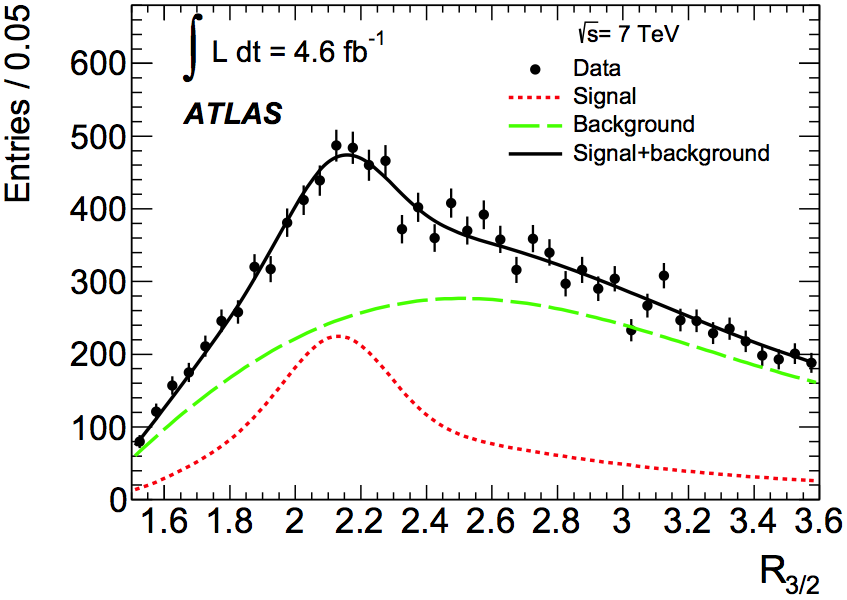}
\put(19,48){(a)}
\end{overpic}
\begin{overpic}[clip,height=5cm]{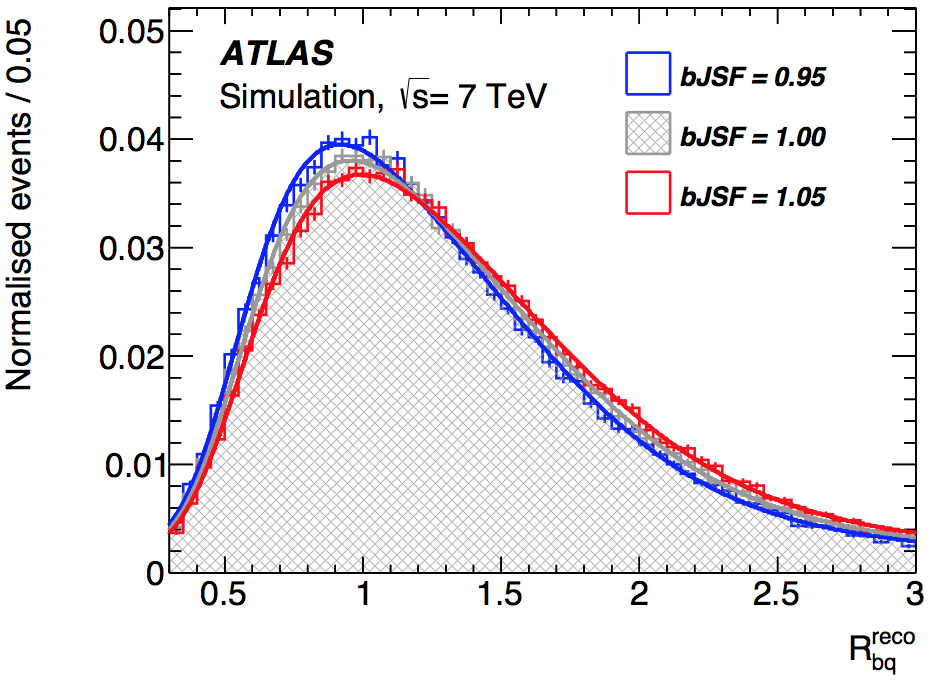}
\put(21,52){(b)}
\end{overpic}
\caption{
\label{fig:jj}
{\bf(a)} The distribution in $R_{3/2}$ in the all-hadronic channel using 4.6~\fb of data at \sseven~\cite{bib:jj}. The signal and background templates are shown for the best-fit values in data.
{\bf(b)} The dependence of the templates in \rbq on the bJSF in the \ljets channel, after reconstruction with a kinematic fit~\cite{bib:lj}.
}
\end{figure}

A measurement of \mt is performed in all-jets events with the highest \ttbar branching ratio using 4.6~\fb of data at \sseven~\cite{bib:jj}. The analysis requires $\geq5$ jets with $\pt>55$~GeV and a $6^{\rm th}$ jet with $\pt>30$~GeV, with two $b$-tags. A veto against isolated leptons ($e$ or $\mu$) and topological selections are imposed. The template method is applied to extract \mt. The dominant systematic uncertainty from the JES is reduced through the usage of the $R_{3/2}=m_{q'\bar q b}/m_{q'\bar q}$ observable, where $m_{q'\bar q b}$ ($m_{q'\bar q}$) is the invariant mass of the jet system matched to the $t\to q'\bar q b$ ($W\to q'\bar q$) decay. The correct assignment of jets to partons is identified through a kinematic likelihood fit. The dominant background from multijet (MJ) production is modelled from data using control regions 
in $\pt^{6^{\rm th}\rm jet}$ and $b$-tag multiplicity. The best fit to data is shown in Fig.~\ref{fig:jj}(a) and results in $\mt=175.1\pm1.4~\stat\pm1.2~\syst$~\GeV. The dominant systematic uncertainties are JES (0.5~GeV), $b$~quark JES (0.6~GeV), alongside PS and hadronisation (0.5~GeV).

\section{Lepton+jets channel} \label{sec:lj}

\begin{figure}
\centering
\begin{overpic}[clip,height=5cm]{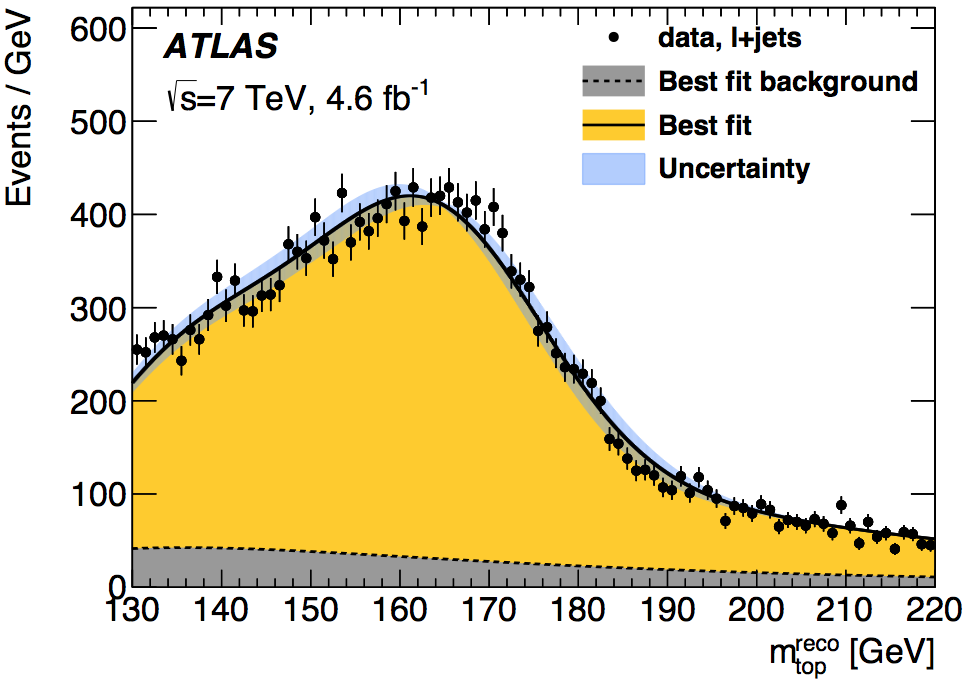}
\put(19,52){(a)}
\end{overpic}
\begin{overpic}[clip,height=5cm]{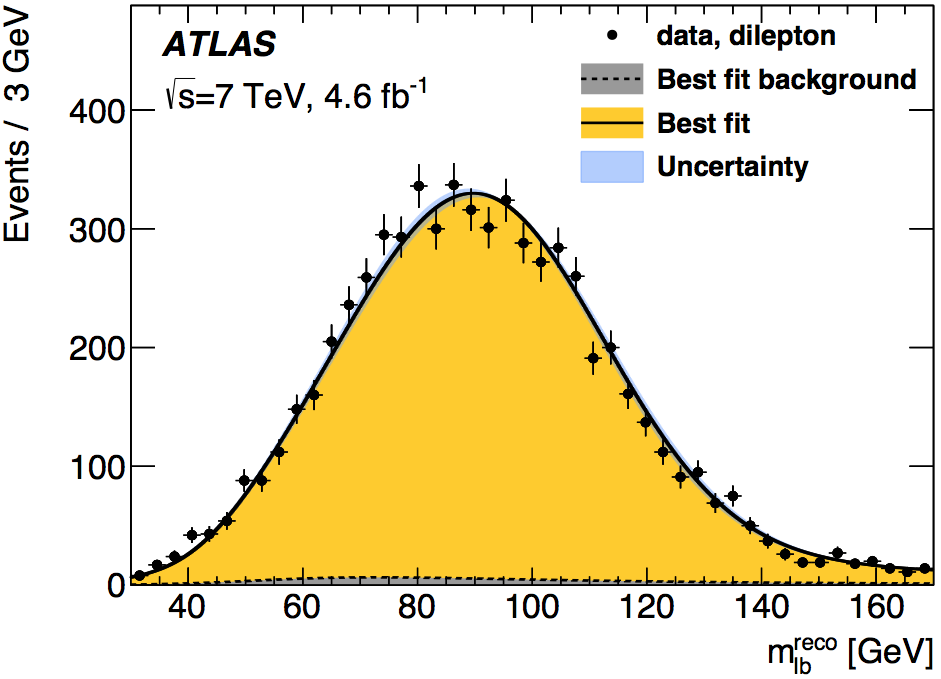}
\put(19,52){(b)}
\end{overpic}
\caption{
\label{fig:lj}
{\bf(a)} The distribution in reconstructed $\mt$ in the \ljets channel using 4.6~\fb of data at \sseven~\cite{bib:lj}. Also shown are the templates for the \ttbar prediction scaled to the best-fit values measured in data, alongside with the background template. 
The shaded band indicates the uncertainty of the template fit.
{\bf(b)}~Same as (a), but for the \dilep channel~\cite{bib:lj}.
}
\end{figure}

ATLAS' most precise single measurement of \mt is done in \ljets events using 4.6~\fb of data at \sseven~\cite{bib:lj}. The selection is similar to that in Sect.~\ref{sec:st}, however, $\geq 4$ jets within $|\eta|<2.5$ are required, with one or two $b$-tags. Like in the all-jets analysis from Sect.~\ref{sec:jj}, the correct assignment of jets to partons is identified through a kinematic fit, and \mt is extracted with a template method. The uncertainty from JES is reduced through a novel approach which simultaneously fits \mt together with the overall JES factor (JSF) and with the ratio of the energy scales for $b$~quark jets to that of $u,d,c,s$ quark and gluon jets (bJSF). The $m_{q'\bar q}$ observable is used to extract JSF, while bJSF is constrained through the $R_{bq}=(\pt^{b}+\pt^{\bar b})/(\pt^{q'}+\pt^{\bar q})$ observable shown in Fig.~\ref{fig:jj}(b) in events with two $b$-tags, and $R_{bq}=2\pt^{b-{\rm tag}}/(\pt^{q'}+\pt^{\bar q})$ for one $b$-tag. Thus, the uncertainty from $b$ quark JES is diminished from $\approx$1~\GeV to 0.06~GeV and other uncertainties are reduced, albeit at the cost of an additional statistical component of 0.67~GeV. The reconstructed \mt for the best fit to data is shown in Fig.~\ref{fig:lj}(a), and the final result reads $\mt=172.33\pm0.75({\rm stat+JSF+bJSF})\pm1.02\syst$~GeV
. The dominant systematic uncertainty remains JES (0.58~GeV), followed by $b$-tagging (0.50~GeV) as the dependence of the $b$-tagging uncertainty on $\pt^{\rm jet}$ impacts the $R_{bq}$ observable.

\section{Dilepton channel} \label{sec:ll}

A measurement of \mt is performed in the \dilep channel using 4.6~\fb of data at \sseven~\cite{bib:lj}. The selection is similar to that in the \ljets analysis from Sect.~\ref{sec:lj}, however, two isolated leptons ($e$ or $\mu$) of opposite charge and $\geq 2$ jets are required. Like for the \mt measurement using single top events from Sect.~\ref{sec:st}, templates in \mlb are used to extract \mt. The best fit to data is shown in Fig.~\ref{fig:lj}(b), and results in $\mt=173.79\pm0.54\stat\pm1.30\syst$~GeV. The dominant systematic uncertainties are the overall JES (0.75~GeV) and the $b$ quark JES (0.68~GeV), followed by modelling of the PS and hadronisation (0.53~GeV) and initial/final state radiation (0.47~GeV).

\section{Combination in lepton+jets and dilepton channels} \label{sec:combi} 

The \mt values measured in the \ljets and \dilep channels from respective Sects.~\ref{sec:lj} and~\ref{sec:ll} are combined using the best linear unbiased estimator, taking into account the signs of the correlation coefficients between the channels for the sources of systematic uncertainty considered~\cite{bib:lj}. With this approach, the \dilep channel can benefit from the constraint on JSF and bJSF in the \ljets channel, which in turn benefits from the smaller statistical uncertainty and the reduced dependence of \mt on the $b$-tagging uncertainty in the \dilep channel. The combination 
results in $\mt=172.99\pm0.48\stat\pm0.78\syst$~GeV, corresponding to a relative uncertainty of 0.53\%.

\section{Top quark mass from $\boldsymbol{\stt}$} \label{sec:mtpole_tt}

\begin{figure}
\centering
\begin{overpic}[clip,height=6cm]{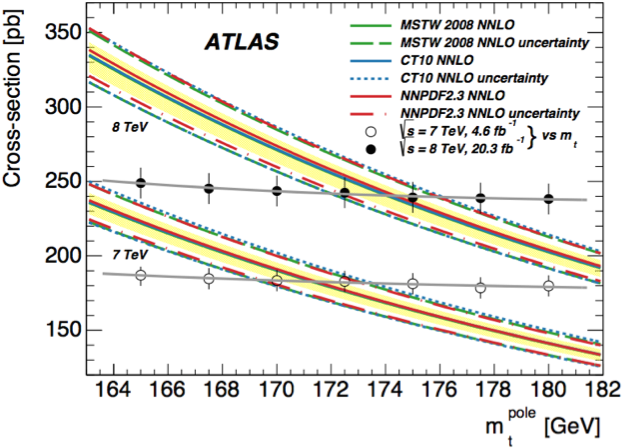}
\put(40,58){(a)}
\end{overpic}
\begin{overpic}[clip,height=6cm]{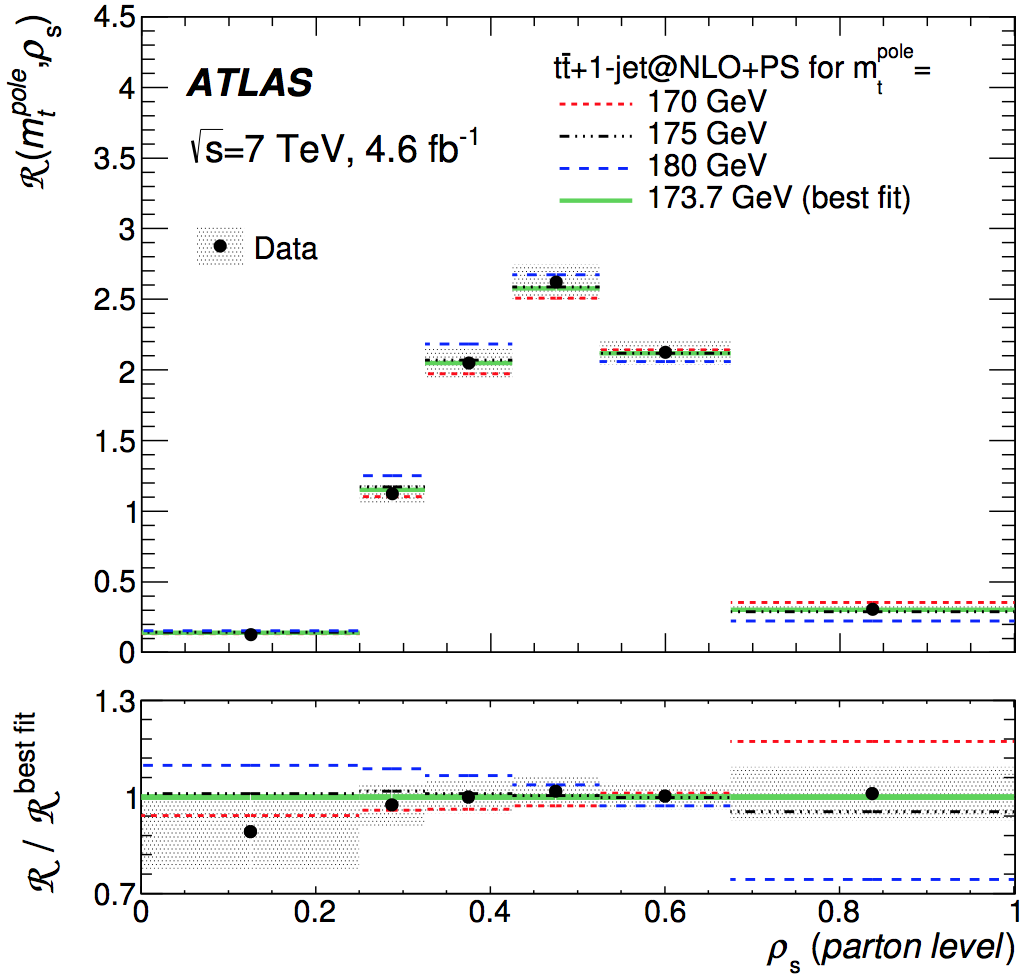}
\put(19,40){(b)}
\end{overpic}
\caption{
\label{fig:mtpole}
{\bf(a)} Predicted NNLO+NNLL \ttbar\ production cross sections at $\sqrt s=7$ and 8~TeV as a function of \mtpole~\cite{bib:mtpole_tt}. The shaded band indicates the QCD scale uncertainty. The measurements of \stt are also shown. 
{\bf(b)}~The distribution $\mathcal R\equiv1/\sigma_{\ttbar+1\rm jet} \cdot \dif \sigma_{\ttbar+1\rm jet}/\dif \rho_s$ at parton level, after corrections for detector, PS, and hadronisation effects, and after background subtraction at \sseven~\cite{bib:mtpole_ttj}. The uncertainty on the data points includes the statistical component only.
}
\end{figure}

The {\em direct} measurements of \mt presented in Sects.~\ref{sec:st}-\ref{sec:combi} provide the most precise experimental determination of \mt. However, they measure the \mt parameter as implemented in MC generators, which is related to the theoretically sound definition in the pole mass scheme \mtpole with an uncertainty of $\leq$1~GeV~\cite{bib:alt}. The world's most precise {\em indirect} measurement of \mtpole from \stt is performed using 4.6~\fb of data at \sseven and 20.3~\fb at \seight~\cite{bib:mtpole_tt}. This analysis exploits the dependence of \stt on \mtpole which is now known with $\approx$3\% precision at NNLO with NNLL corrections~\cite{bib:xsec_tt_nnlo}, and extracts \mtpole using the measurement of \stt in $e^\pm\mu^\mp$ final states. The input measurement of \stt achieves an unprecedented relative uncertainty of $\approx$4\% through widely applying data-driven techniques and extracting \stt simultaneously with the $b$-tagging efficiency, which would otherwise represent the largest source of experimental uncertainty. 
The resulting distributions of extracted $\stt(\mtpole)$ for \sseven and \seight are given in Fig.~\ref{fig:mtpole}(a), and show a very small dependence on \mtpole due to  minimal use of MC simulations in the analysis. The \mtpole parameter is extracted using a maximal likelihood fit of the experimental result and the theoretical calculations~\cite{bib:xsec_tt_nnlo}. The final results are given by $\mtpole=171.4\pm2.6~$GeV at \sseven, $\mtpole=174.1\pm2.6~$GeV at \seight, and $\mtpole=172.9^{+2.5}_{-2.6}~$GeV in combination. 
The dominant sources of systematic uncertainty on the experimental side are the luminosity~(0.9~GeV) and the beam energy~(0.6~GeV), while the choice of parton distribution functions and $\alpha_S$~(1.7~GeV) and the choice of factorisation and renormalisation scales~(1.2~GeV) dominate the theory side.

\section{Top quark mass from $\boldsymbol{\sttj}$} 

The world's first measurement of \mtpole from the production cross section of a \ttbar system in association with a jet \sttj is performed in the \ljets channel using 4.6~\fb of data at \sseven~\cite{bib:mtpole_ttj}. The sensitivity to \mtpole is established through the radiation rate of a high-\pt gluon off the \ttbar system, which is proportional to \mtpole. More precisely, the differential production cross section $\mathcal R(\mtpole,\rho_s)\equiv1/\sigma_{\ttbar+1\rm jet} \cdot \dif \sigma_{\ttbar+1\rm jet}/\dif \rho_s$ is compared to NLO calculations~\cite{bib:mtpole_ttj_xsec}, where $\rho_s \equiv 2m_0/\sqrt{s_{\ttbar+1{\rm jet}}}$, and the arbitrary constant $m_0$ is set to 170~GeV in this analysis. The selection is similar to the \ljets analysis in Sect.~\ref{sec:lj}, however, two $b$-tagged jets are required in each event. The reconstruction of the \ttbar system is performed through a $\chi^2$ kinematic fit. To reduce the total uncertainty, $\pt>50~\GeV$ is required for the extra jet. The distribution in $\rho_s$ is corrected for detector, PS, hadronisation effects, and the presence of background. The resulting distribution at parton level is given in Fig.~\ref{fig:mtpole}(b). The extraction of \mtpole is performed through a $\chi^2$ fit of the observed and expected values in each bin of the $\rho_s$ distribution for the \mtpole parameter, and yields $\mtpole=173.1\pm1.50\stat\pm1.43\syst^{+0.93}_{-0.49}$~GeV. 
The dominant experimental systematic uncertainties are the contributions from the modelling of initial and final state radiation~(0.72~GeV) and from the JES~(0.94~\GeV). The theory uncertainty is dominated by the choice of factorisation and renormalisation scales.

\section{Conclusions}
I presented recent measurements of the top quark mass, a fundamental parameter of the SM, by the ATLAS Collaboration. ATLAS' most precise determination of $\mt=172.99\pm0.48\stat\pm0.78\syst$~GeV is attained in the combination of the \ljets and \dilep channels. Innovative techniques have been put in place, like the first measurement of \mt in single top events, the first {\em in situ} calib\-ration of the $b$ quark JES, and the world's most precise measurements of \mtpole from \stt and \sttj. 

\section*{Acknowledgments}
I would like to thank my colleagues from the ATLAS experiment for their help in preparing this article, the staffs at CERN and collaborating institutions, as well as the ATLAS funding agencies.

\end{document}